\documentclass{iaus}
\usepackage{graphicx}

\newcommand{\ltsima}{$\; \buildrel < \over \sim \;$}
\newcommand{\simlt}{\lower.5ex\hbox{\ltsima}}
\newcommand{\gtsima}{$\; \buildrel > \over \sim \;$}
\newcommand{\simgt}{\lower.5ex\hbox{\gtsima}}

\title[Mass assembly from chemical abundances] %% give here short title %%
{How  galaxies form: Mass assembly from chemical abundances in the era of large surveys}

\author[Rosemary F.G.~Wyse]   %% give here short author list %%
{Rosemary F.G.~Wyse}
\affiliation{Department of Physics \& Astronomy, Johns Hopkins University, \\ Baltimore, MD 21218, USA  \\ email: {\tt wyse@pha.jhu.edu}}

\pubyear{2009}
\volume{265}  %% insert here IAU Symposium No.
\pagerange{119--126}
\jname{Chemical Abundances in the Universe: Connecting First Stars to Planets}
\editors{K. Cunha, M. Spite \& B. Barbuy, eds.}
\begin{document}

\maketitle

\begin{abstract}

The chemical abundances in the atmosphere of a star provide unique
information about the gas from which that star formed, and, modulo processes that are not important for the vast majority of stars, such as 
mass transfer in close binary systems, are conserved through a star's
life.  Correlations between chemistry and kinematics have been used
for decades to trace dynamical evolution of the Milky Way Galaxy. I
discuss how it should be possible to refine and extend such analyses, provided planned large-scale deep imaging surveys have matched spectroscopic surveys.

\keywords{stars: abundances, stars: kinematics, Galaxy: abundances, Galaxy: evolution, Galaxy: formation, Galaxy: stellar content, (galaxies:) Local Group; 
cosmology: dark matter}
%% add here a maximum of 10 keywords, to be taken form the file <Keywords.txt>
\end{abstract}

\firstsection % if your document starts with a section,
              % remove some space above using this command.
\section{Near-Field Cosmology}

Analysis of the properties of old stars in our own Milky Way galaxy,
and in nearby galaxies, can be used to infer physical conditions in
the early Universe, when those stars formed. Many of the properties of
stars are approximately conserved over long timescales, even in the
hierarchical merging scenario for galaxy formation. We thus can trace
the evolution of any one individual galaxy through time, from the
`fossil record' imprinted in the motions and chemical abundances of
its stars -- for example, the Milky Way, with current capabilities,
and even beyond the Local Group with planned ELTs.  This complements
beautifully the one-age snapshots of many galaxies that are available
from direct studies of high-redshift objects.  Further, when one
observes high-redshift systems, one sees the total sum of all their
stars together, and the interpretation of this `integrated light' is
subject to several degeneracies (e.g.~between stellar age and
metallicity, or between star-formation rate and stellar Initial Mass
Function) that limit the robustness of conclusions that may be
drawn. In contrast, when individual stars are observed, these
degeneracies can be broken. Modeling of stellar motions further can
map dark matter distributions in three-dimensions, critical
information for discriminating different dark matter candidates. 

A major prediction of $\Lambda CDM$ cosmology is that much of the
stellar populations of present-day large galaxies, like the Milky Way,
formed in systems of much lower mass. The bulge and stellar halo are
expected to be built-up during mergers, with predominantly
dissipationless mergers dominating for the halo, and both
dissipational and dissipationless mergers contributing to the bulge.
Stellar disks can be thickened by mergers, and satellites on the right
orbits can even contribute stars (and dark matter!) directly to both
the thick and thin disks. 

Spectroscopy provides much of the necessary astrophysics to test
these, and other, models, and, ideally, together with astrometric
imaging surveys, allows the acquisition and analysis of full
three-dimensional space motions, elemental abundances and spatial
distributions of large samples of stars. Spectra provide the
line-of-sight velocities, and the stellar parameters -- effective
temperature, gravity and chemical abundances -- with an accuracy that
depends on wavelength region, signal-to-noise and spectral resolution.
The panoramic imaging surveys being planned and implemented --
including Gaia -- need matching spectroscopic surveys to fulfil their
full potential in deciphering the nearby Universe.
The physics of galaxy formation and evolution that can then be
addressed includes deciphering the spatially resolved star-formation
histories of galaxies, the mass-assembly histories -- which may be
quite different from the star-formation histories, and their links to
growth of a central black hole. Further, the role and form of
`feedback' that is necessary in CDM theories to regulate star
formation must be consistent with the fossil record in the stars that
form.  The nature of dark matter is constrained not only by the halo
potential well shape, but also by the merger histories of galaxies,
the effects of which again are written in the properties of the stars
both within -- and outwith -- galaxies.

\section{Scientific Requirements and Capabilities}

 	The combination of an estimate of overall metallicity (usually
calibrated onto iron) to 0.2~dex or better, plus radial velocities to
10-20~km/s, photometric distances (to 20-30\%), and ideally also age
and proper motions from available photometry, can be used to assign
probabilistically a star to its parent stellar component (thin disk,
thick disk, bulge, bar, halo, satellite system, plus the unexpected).
Of course there will always be ambiguities since different populations
have distributions that overlap, exacerbated if simple Gaussians are
assumed for the underlying parent distributions.  Large sample sizes
allow the definition of populations beyond mean properties, which is
important since much physics resides in the detailed shape of
distributions: mean metallicity tracks depth of potential well, while
the shapes of the wings are more sensitive to pre-enrichment, and gas
flows etc. Quantification of stellar properties at this level suffice
to define the overall chemical evolution and star-formation history of
a given (portion of!) a stellar population, and can be used to
constrain the merger history of the parent galaxy, and map dark matter
potential wells of systems of mass of the Milky Way.  All of this can
be achieved by medium-resolution optical/NIR spectroscopy (${\cal {R}}
\sim 5,000$).

Precise and accurate elemental abundances require high S/N, high
spectral resolution data (${\cal{R}} \sim 50,000$).  Such abundance
data contain significantly more information than does overall
metallicity, since the latter integrates over star-formation history,
while the former retains such information, through the stellar-mass
and temporal dependence of chemical yields: to first order, different
elements are produced by different mass stars on different
timescales. Thus the relative contributions of core-collapse (Type~II)
supernovae, with significant products of pre-explosion, steady-state
nucleosynthesis and of white-dwarf explosive nucleosynthesis (Type~Ia)
supernovae can be estimated from the patterns of element ratios.
Isolating stars that were predominantly (pre)enriched by only
core-collapse SNe allows one to search for variations in the
massive-star IMF (e.g.~Wyse \& Gilmore 1992; Nissen et al.~1994) --
and the evidence from such resolved-star studies is that, while there
is tentative evidence in the most metal-poor stars, [Fe/H$ \simlt -3$,
for enrichment by a small number of SNe and thus incomplete mixing of
metals in the interstellar-medium, there is surprising uniformity of
the elemental abundances, implying uniformity of the massive-star
IMF. The distribution of delay times (after formation of the
progenitor of the white dwarf that explodes) prior to Type Ia
supernovae is model dependent, and plausibly a mix of
double-degenerate and single-degenerate channels could well
contribute (e.g.~Matteucci et al.~2009). Clearly the minimum delay time equals  the time to form
the white dwarf descendent of the most-massive progenitor star, $\sim
8-10$M$_\odot$. The maximum delay time depends on the mass ratio and
orbital parameters of the binary system, and in principle is a Hubble
time. Chemical evolution models of a self-enriching system typically
predict that the signature of significant contributions from Type~Ia
-- a characteristic decrease in [$\alpha$/Fe] -- does not appear until
$\sim 1$~Gyr after the onset of star-formation. Identifying the iron
abundance at which one sees the downturn allows a time to be
associated with that level of enrichment.  This is of obvious
importance for modelling star-formation histories.

\section{Applications}

\subsection{The smooth stellar halo}

Photometric techniques can obviously be applied to fainter targets
than can spectroscopic techniques, and prior to multi-object
spectrographs, had a significant multiplex advantage also.  In terms
of chemical abundances, the line-blanketing in the U-band offers the
most powerful approach, utilising a comparison of $(U-B)$ in a given
star to that of a star of the same $(B-V)$ in a cluster of known
metallicity (e.g.~the Hyades). Sandage (1969) developed the
calibration of $\delta(U-B)_{0.6}$, using a normalisation to a fixed
value of $(B-V)=0.6$ to take out the temperature sensitivity. This is
valid only for a limited range of effective temperature, essentially
F/G main sequence stars, and it saturates at low metallicities, $\sim
-1.5$~dex (e.g.~Carney 1979, his Fig.~3), understood in terms of the opacity from metals becoming comparable to that from helium, so that reducing the metallicity results in little further reduction in overall opacity.  Within the range of
validity, it provides metallicities accurate to $\sim 0.2$~dex, for
1\% photometry.  The advent of wide-field, deep photometry from the
Sloan Digital Sky Survey has opened the opportunity to develop and
apply such techniques using the photometry in the SDSS filters.  The
result of such an analysis, based on $ugr$ photometry, calibrated
using the SDSS stellar parameters from the spectroscopic pipeline, is
vividly illustrated in Ivezic et al.~(2008): the halo, out
to $\sim 10$~kpc from the Sun, is remarkably uniform in metallicity,
with no gradients or scatter about the mean value of [Fe/H] $=
-1.46$~dex -- their quoted deviation about the median of $\sim
-1.5$~dex is less than 0.05~dex.  Even more distant large-scale
overdensities, such as the Virgo overdensity at 10--20~kpc, have the same mean iron abundance as the rest of the field halo, based on photometric
metallicities from $gri$ photometry (An et al.~09).  This is somewhat
unexpected, implying very well-mixed gas across the star-forming
regions that created the halo. The simulations of Johnston et
al.~(2008) suggest that inhomogeneities should have been
observed. Spectroscopic confirmation of the `spatially invariant
Gaussian' metallicity distribution of the (inner?) halo would provide
rather stringent constraints on theories.

Spectroscopy also provides  line-of-sight velocities, allowing the
identification of kinematically defined subsystems in addition to
metallicity distributions.  The spectroscopic component of the initial
SDSS galaxy survey utilised Galactic stars for spectrophotometric
calibration and removal of telluric features. These stars can be used
to investigate the Galactic stellar halo and thick disk, although
their non-standard selection functions preclude the detailed analysis
called for above (dedicated stellar surveys should provide that).
Carollo et al.~(2008, 2009) demonstrated that this calibration-star
sample is best-fit by a two-component smooth halo, with the components
distinguished by both kinematics and metallicity, and also radial
distribution -- the outer halo is more metal-poor and in retrograde
rotation, compared with the non-rotating inner halo. This is
reminiscent of the seminal results of Searle \& Zinn (1978), based on globular clusters, with much
later interpretation in terms of hierarchical clustering $\Lambda$CDM
models of Galaxy formation.

\subsection{Elemental abundances and kinematics}

The combination of detailed elemental abundances with kinematic and/or
spatial phase space information is immensely powerful.  This was
beautifully illustrated by the compilation of elemental abundance data
for stars in the Milky Way and its satellite galaxies that was shown
at this Symposium by Andreas Koch (see his contribution for details,
also Koch 2009; Tolstoy et al.~2009; Geisler et al.~2007).  With no coding to indicate parent galaxy,
the plot of [Fe/H] against [Ca/Fe] is largely a scatter plot. However,
once stars are coded by parent galaxy, it is evident that each galaxy
has its own locus in this plane. This can be understood in terms of
the wide diversity of star-formation histories from satellite to
satellite (e.g.~Hernandez, Gilmore \& Valls-Gabaud, 2000; Orban et
al.~2008), and the relatively long duration of star formation in the
vast majority of satellites, so that Type~Ia supernovae are important
contributors to the chemical enrichment. A consequence is that at a
given [Fe/H], [$\alpha$/Fe] is typically lower in stars in dwarf
galaxies than in the field halo, reflecting the longer duration of
star formation in the former compared to the latter. As noted above,
an invariant IMF leads to invariant elemental abundance ratios (with
some scatter if mixing and/or sampling of the mass function is
incomplete) if only core-collapse supernovae contribute, with nothing
to distinguish the parent system. The accretion and merging of
independent star-forming satellites, with typical extended
star-formation histories, may be expected to produce structure in the
element abundance patterns of stars in the host larger galaxy.
Elemental abundances can thus be used to define and map substructure,
and place constraints on the minor-merger history of the larger
galaxy. It may be noted that at the lowest iron abundances in each
satellite, there should be stars with enhanced [$\alpha$/Fe],
consistent with the field halo, being the first stars to form within that system, and
this is now being observed (e.g.~Cohen \& Huang 2009; Frebel et al.~2009; Norris et
al.~2010a).

\begin{figure}[h]
% \vspace*{-2.0 cm}
\begin{center}
 \includegraphics[angle=270,width=4in]{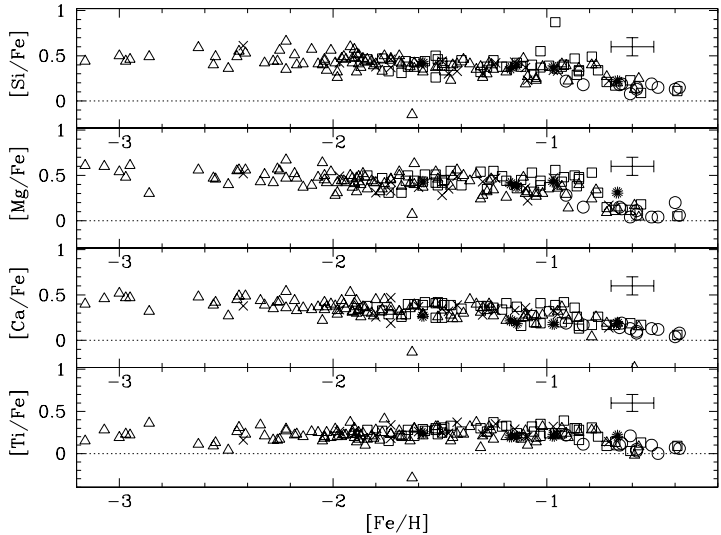}
% \vspace*{-1.0 cm}
 \caption{Elemental abundance ratios for metal-poor disk and halo 
 stars, initially selected from the medium-resolution RAVE spectroscopic survey.  The stars have been assigned to a given population based on a combination of kinematic and positional criteria. Open  circles represent thin-disk stars, open  squares 
thick-disk stars, open triangles are halo stars, asterisks are thin/thick disk, and crosses are thick disk/halo.  The thick and thin disks clearly extend to low iron abundances, and at these low abundances all populations have the same (enhanced) 
value of [$\alpha$/Fe], unlike the bulk of stars in
 satellite galaxies.  The most robust conclusion is that the same massive-star IMF pre-enriched halo, thick disk plus thin disk, and that for all three this IMF was well-sampled, and there was also good mixing.  } \label{fig4}
\end{center}
\end{figure}

For the Milky Way at least (at present) one can go further than simply
coding stars by their parent galaxy, by (probabilistically) assigning
stars to individual stellar components of the parent galaxy.  This is
illustrated in Figure~1, based on Ruchti et al.~(2010; part of Greg
Ruchti's PhD thesis at JHU).  The main scientific thrust of this work
is to identify the low-metallicity stars of the thick and thin disks,
as probes for the early evolution of these components, and compare the
derived elemental abundances with the predictions of theories. For
example, build-up of the disks by late accretion, directly into the disks, of stars from merging
satellite galaxies (e.g.~Abadi et al.~2003)
would predict metal-poor stars with low values of [$\alpha$/Fe].  The
RAVE spectroscopic survey (Steinmetz et al.~2006) provides
moderate-resolution spectra of bright stars, allowing estimates of
overall metallicity and distances.  The stars are sufficiently
bright that estimates of proper motion are available, giving full
space motions.  From this, one can select a sample of candidate
metal-poor stars with disk kinematics, for follow-up echelle
spectroscopy (the RAVE stars are ideal targets for 4-8m class telescopes).
Our present sample is shown in Fig.~1, where elemental abundances for some 170 
stars are shown, coded by the stellar component to which the star most probably belongs.

There are several conclusions from this figure.  The disks extend to
low metallicity, [Fe/H] $\sim -1$~dex for the thin disk and $\sim
-1.75$~dex for the thick disk. There is remarkably little scatter in the patterns of the elemental abundance ratios, and
at low iron abundances, all components show the same enhanced
[$\alpha$/Fe] (unlike stars in satellite galaxies).  This implies pre-enrichment by the same underlying massive-star IMF, with
rapid (short-duration) star formation taking place in regions that are
well-mixed, and of high enough mass that the IMF is well-sampled. There is no evidence supporting a variation of the massive-star IMF. Neither is there any evidence to support late accretion of stars from satellite galaxies directly into the disks, as this would be expected to create metal-poor disk stars with low [$\alpha$/Fe] which is not observed. 

It is also apparent that  the
stellar halo could form by the merging of stars formed in any
system(s) in which star-formation is short-lived, and chemical
evolution is truncated/inefficient so that the mean metallicity is
kept low; these systems could be star clusters, galaxies, or transient
structures in which stars form during accretion of gas into the Galaxy; identifying stars with enhanced [$\alpha$/Fe] is not sufficient to identify halo progenitors. The addition of	complementary, independent age information -- that the bulk of field halo stars are {\it old\/} --  allows us to constrain further possible progenitors, and rules out late accretion of typical luminous dwarf galaxies (e.g.~Unavane, Wyse \& Gilmore 1996).

The sample of Fig.~1 was defined using a rather complex selection
function and thus understanding the relative numbers of thick and thin
disk stars requires modelling (underway). As discussed in the
contributions of Bensby and by Reddy to these proceedings (see also
Bensby et al.~2007a), there is, in some samples of local disk
stars,`kinematic confusion', in that stars selected on the basis of
their kinematics to be thick-disk members can follow the elemental
abundance trend of the thin disk, as opposed to that of the thick
disk. Using age information as part of the criteria used to assign a
given star to a given component appears to minimise the confusion
(Bensby \& Feltzing 2009) -- the older stars follow the thick disk
elemental abundance trends. However, we know that the velocity
distribution function in the local thin disk (probed by the Hipparcos satellite)
contains `moving groups' and is non-Gaussian, in at least the radial
and azimuthal motions (e.g.~Dehnen 1998; Famaey et
al.~2005). Resonances in the disk plane due to transient spiral arms
and the Galactic bar provide plausible explanations for this kinematic
structure (e.g.~Dehnen 2000; de Simone et al.~2004; Bensby et
al.~2007b). The detected dynamical substructures have cold vertical
velocity dispersions, lower than typical for the thin disk (Famaey et
al.~2005). The adoption of Gaussian velocity distributions for the
underlying Galactic components when making assignments of stars to a
given component is clearly an over-simplification, one that should --
and could -- be abandoned.  Large, unbiased surveys such as RAVE can
be used to determine the actual velocity distributions, for use instead. 

Dynamical interactions between stars and transient spiral arms can
also lead to radial migration of resonant stars (e.g.~Sellwood \&
Binney, 2002; Ro\v{s}kar et al.~2008a,b), which in turn can be
expected to imprint a signature in the elemental abundance pattern of
local stars, reflecting the variation in star-formation history and
chemical evolution across the region of migration (e.g.~Sch\"onrich \&
Binney 2009), which may even have been detected (Haywood 2008).
However, as noted in Sch\"onrich \& Binney (2009), the efficiency of migration should be a function of several parameters such as velocity dispersion, and theoretical understanding of this is incomplete.

\subsection{Ultra-faint satellite galaxies}

Present samples of stars with elemental abundances from high-spectral
resolution observations are restricted to $\sim $ a thousand bright
field stars (e.g.~compliation of Roederer 2009), and hundreds of stars
in satellite galaxies (e.g.~Koch, this volume); more data are clearly
needed!  The `ultra-faint' satellite dwarf galaxies are clearly very
interesting for study, with intriguing results even from very small
samples. Our results for radial-velocity members of Bo\"{o}tes~I
(discovered by Belokurov et al.~2006; luminosity L$_V \sim 3 \times 10^4$L$_\odot$, distance $\sim 60$~kpc; Martin, de Jong \& Rix
2008) and of Segue~1 (discovered by Belokurov et al.~2007; L$_V \sim 335$L$_\odot$, distance $\sim 25$~kpc; Martin et al.~2008) are
shown in Fig.~2 (see also Norris et al.~2008; Norris et
al.~2010b). Here I show derived iron and carbon abundances, based on
intermediate-resolution spectra (from the 2-degree field of view,
multi-object, fibre-fed spectrograph AAOmega on the Anglo-Australian
Telescope) covering the blue Ca~II~K-line (from which the iron
abundance is derived) and the CH G-band (synthesis of which is the basis for the carbon
abundance). Thus the iron abundances are on the same scale
as many of the surveys of metal-poor field halo stars (see Beers \&
Christlieb 2005). 

\begin{figure}[h]
\vskip -2.2 truein
% \vspace*{-2.0 cm}
\begin{center}
 \includegraphics[angle=270,width=5.5in]{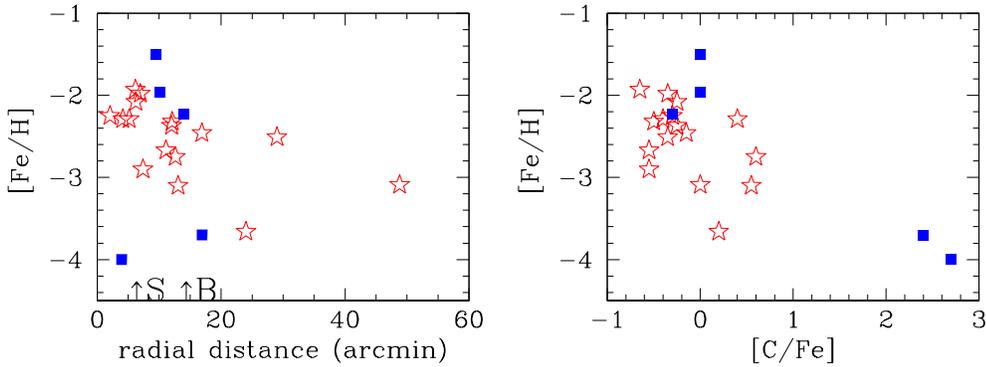}
% \vspace*{-1.0 cm}
 \caption{Iron and carbon abundances for stars that are radial-velocity members of the Milky Way satellite systems Bo\"{o}tes~I (open red star symbols) and Segue~1 (filled blue square symbols). The mean level of chemical enrichment is low, and there is a large dispersion within each system. Similarly to the field halo, the most iron-poor stars are carbon-rich.}
\label{fig2}
\end{center}
\end{figure}

The left panel of Fig.~2 shows the iron abundances
as a function of the projected radial distance of a given star from
the centre of its parent galaxy, with the filled (blue) squares
denoting stars that are radial-velocity members of Segue~1, and the
open (red) star symbols denoting radial-velocity members of Boo~I.
The vertical arrows along the $x$-axis mark the half-light radii of
these two systems (`S' for Segue~1 and `B' for Boo~I).  The wide
field-of-view of AAOmega allows efficient mapping of these very sparse, nearby 
systems (and is critical given the apparent contamination in these lines-of-sight by stars from the Sagittarius dSph; see Niederste-Ostholt et al.~2009). Three points are immediately obvious: the mean level of
enrichment is low, with a fraction of `extremely metal-poor' stars
(those with [Fe/H] $ < -3.0$~dex; see Table~1 of Beers \& Christlieb
2005); there are members well beyond the nominal half-light radius;
and there is a large internal metallicity dispersion in each system,
suggestive of self-enrichment. The right panel of Fig.~2 shows the
carbon-to-iron ratio in these stars; here the striking result is the
very high values of [C/Fe] for the two most metal-poor stars in
Segue~1 (indeed, the values derived for [Fe/H] for these two C-rich
stars are rather uncertain given the weakness, and probable CH
contamination, of the CaII~K line in their spectra). As discussed by
many authors (e.g.~Norris et al.~2007), the fraction of carbon-rich
stars in the field halo increases as the iron abundance decreases, and
this has significant implications for both early supernovae and
cooling mechanisms for the early interstellar medium.  The mass
densities in the inner regions may be estimated in the more luminous
dSph, for which large enough samples of stars with line-of-sight
velocities over the radial extent of the system can be obtained,
enabling mass-profile fitting (e.g.~Gilmore et al.~2007).  These inner densities ($\rho_{DM} \sim 0.1$~M$_\odot$/pc$^3$) are such that the implied redshift of collapse is $\sim 15$, around the expected epoch of reionization. Constraints on the star-formation process in these systems is of obvious interest. Follow-up high-resolution observations are underway at the VLT.

\subsection{The Inner Galaxy: built-up by mergers, a starburst, disk instability?}

The chemical abundance distribution of the Galactic bulge has been
probed, using spectroscopy of giant stars, in several low-extinction
optical windows (e.g.~Rich, 1988; Ibata \& Gilmore 1995a,b; Sadler,
Rich \& Terndrup 1996; Zoccali et al.~2003), with a recent extension
to elemental abundances in a few lines-of-sight (e.g.~Zoccali et
al.~2008; Fulbright, McWilliam \& Rich 2007; Ryde et al.~2009).  The
general concensus is that the central bulge has a high mean iron
abundance, $\sim -0.25$~dex (similar to that of the local thin disk),
with a suggestion of a radial gradient.  The elemental abundances show
enhanced [$\alpha$/Fe] for the bulk of the stars, pointing to a
short-lived burst of star formation, with the high overall abundance
requiring that this occur in a deep potential well, so that
supernova-ejecta are retained. At abundances around the solar value,
there appears to be a turn-down in the element ratios, which may be
explained in terms of metallicity-dependent yields of core-collapse
supernovae (Fulbright et al.~2007). The presence of metallicity gradients suggests dissipation
during star formation.  Again, complementary age information is
available, from deep color-magnitude
data and point to an old age for the dominant population, $\sim
10-12$~Gyr (e.g.~Clarkson et al.~2008 from optical HST data, and van Loon et al.~2003 from IR data from the ISO satellite),
strengthening the arguments for a short-lived duration of star
formation. Slow build-up of the bulge is not favoured from the chemistry. 

There are several caveats and limitations of the data, however.  There
remains a need to map the bulge/bar and inner disk, to understand
better how the bulge connects to each of the disk and halo. The
ubiquitous intermediate-age tracers, such as AGB stars (e.g.~van Loon
et al.~2003; Uttenthaler et al.~2008) and OH/IR stars (e.g.~van der
Veen \& Habing 1990) need to be placed in context. The planned surveys
APOGEE and HERMES should certainly help.

Evolved stars are clearly the only feasible tracers for which large
samples are available. Elemental abundances of unevolved bulge stars
are possible to obtain if one can exploit fortuitous magnification by
a microlensing event.  This is clearly going to be a rare occurrence,
leading to small samples.  However, the first results are intriguing
and show a very high mean metallicity, apparently not consistent with
the previous results from giant stars (Johnson et al.~2007, 2008; Cohen et
al.~2008, 2009; Bensby et al.~2009), leading to the suggestion that
the giant phase of stellar evolution does not yield unbiased tracers,
but lacks the highest metallicity stars. The statistics of microlensing favour a location at the distance of the bulge for the source stars, but there remains the possibility that they are in the inner disk, rather than the bulge.  Again, we need good mapping of the inner disk -- bulge transition.

\section{Concluding remarks}

Large scale spectroscopic surveys of Galactic stars - and of stars in
Local Group galaxies - are clearly required for both kinematic/dynamic
analyses and chemical abundance determinations.  While several
moderate-size surveys are underway or being planned, such as SEGUE-II,
RAVE, HERMES, APOGEE, these are not sufficient for global
understanding, nor will they match the planned deep imaging 
surveys. Spectroscopic surveys at both moderate- and high-spectral
resolution are needed, and the large samples for statistical analyses call for wide-field
multi-object spectrographs. 

\acknowledgements

I am very grateful to the organizers, to the Brazilian Astronomical Society (SAB) and to the IAU for their financial support, enabling me to participate.

\end{document}